\documentclass[preprint,5p,sort&compress]{elsarticle}

\usepackage[toc,page]{appendix}
\usepackage{amssymb}
\usepackage{amsmath}
\usepackage{hyperref}
\usepackage{xcolor}
\usepackage{listings}

\usepackage{placeins} 
\usepackage{multirow} 
\usepackage{subcaption}
\RequirePackage[capitalise]{cleveref}
\usepackage{mine}

\newcounter{bla}

\journal{Computer Physics Communications}

\usepackage{tikz}
\usetikzlibrary{decorations.pathmorphing,decorations.markings}
\usetikzlibrary{decorations.pathreplacing}
\usetikzlibrary{positioning, calc, shapes}

\usetikzlibrary{shapes.geometric, arrows, calc}
\tikzstyle{tool} = [rectangle, rounded corners, minimum width=1cm, text
					centered, draw=black, fill=red!30]
\tikzstyle{func} = [rectangle, dashed, minimum width=1cm, minimum height=0.6cm,
					text centered, draw=black, fill=green!30]
\tikzstyle{class} = [rectangle, rounded corners, minimum height=0.6cm, minimum
					 width=1cm,text centered, draw=black, fill=blue!30]
\tikzstyle{block} = [rectangle, dashed, minimum width=1cm, minimum height=0.6cm,
					 text centered, draw=black, fill=white]

\tikzstyle{arrow} = [thick,->,>=stealth]
\tikzstyle{line} = [thick,-,>=stealth]

\RequirePackage[status=draft,inline,nomargin,marginclue]{fixme}
\fxusetheme{color}
\FXRegisterAuthor{fh}{afh}{FH} 
\FXRegisterAuthor{cs}{acs}{CS} 
\FXRegisterAuthor{jcm}{ajcm}{JCM} 

\begin{document}

\begin{frontmatter}



\title{Pineline: Industrialization of High-Energy Theory Predictions}


\author[a]{Andrea Barontini}
\ead{andrea.barontini@mi.infn.it}
\author[a]{Alessandro Candido}
\ead{alessandro.candido@mi.infn.it}
\author[b]{Juan M. Cruz-Martinez\corref{author}}
\ead{juan.cruz.martinez@cern.ch}
\author[a]{Felix Hekhorn}
\ead{felix.hekhorn@mi.infn.it}
\author[c]{Christopher Schwan}
\ead{christopher.schwan@physik.uni-wuerzburg.de}

\cortext[author] {Corresponding author.\\\indent\indent\textit{Preprint numbers:} TIF-UNIMI-2023-4, CERN-TH-2023-021}
\address[a]{%
  TIF Lab, Dipartimento di Fisica, Universit\`a degli Studi di Milano and INFN
  Sezione di Milano, \\
  Via Celoria 16, 20133, Milano, Italy
}
\address[b]{CERN, Theoretical Physics Department, CH-1211 Geneva 23, Switzerland}
\address[c]{%
  Universit\"at W\"urzburg, Institut f\"ur Theoretische Physik und
  Astrophysik, 97074 W\"urzburg, Germany
}

\begin{abstract}
  We present a collection of tools efficiently automating the computation of large sets of theory predictions for high-energy physics.
  Calculating predictions for different processes often require dedicated programs.
  These programs, however, accept inputs and produce outputs that are usually very different from each other.
  The industrialization of theory predictions is achieved by a framework which harmonizes inputs (runcard, parameter settings), standardizes outputs
  (in the form of grids), produces reusable intermediate objects, and carefully tracks all meta data required to
  reproduce the computation. Parameter searches and fitting of non-perturbative objects are exemplary use cases that require a full
  or partial re-computation of theory predictions and will thus benefit of such a toolset. As an example application we present a study of the impact of
  replacing NNLO QCD K-factors with the exact NNLO predictions in a PDF fit.
\end{abstract}


\end{frontmatter}

\noindent
{\bf PROGRAM SUMMARY}
\\

\begin{small}
\noindent
\textit{Program Title:} \texttt{pineline} \\
\\
\textit{Program URL:} \url{https://nnpdf.github.io/pineline/}\\
\\
\textit{Licensing provisions:} GPLv3 \\
\\
\textit{Programming language:} Python, Rust \\
\\
\textit{Nature of the problem:} The computation of theoretical quantities in particle physics
often involves computationally-intensive tasks such as the calculation of differential cross sections
in a systematic and reproducible way. Different groups often use different conventions and
choices which makes tasks such as the fitting of physical parameters or quantities very computationally challenging and hard to reliably reproduce.\\
\\
\textit{Solution method:} We create a pipeline of tools such that a user can define an observable
and a theory framework and obtain a final object, containing all relevant theoretical information.
Such objects can be then used in a variety of interchangeable ways (fitting, analysis, experimental comparisons).\\

\end{small}


\section{Introduction and motivation}
\label{sec:intro}

Modern particle physics phenomenology is increasingly reliant on complex theoretical calculations whose accuracy needs to match very precise measurements, chiefly the ones from experiments at the Large Hadron Collider (LHC)~\cite{Amoroso:2022eow}.
An increase in accuracy in those predictions is associated to the computation of higher orders in the strong and/or electroweak couplings for partonic cross sections, and usually performed by numerical programs, which we will call \emph{generators} throughout this paper.
Since the computations are very demanding in runtime, memory and storage, these generators are usually optimized for and can only calculate a small set of observables, and, furthermore, they often use different conventions and strategies.
Being able to generate, store and exchange predictions in well-suited formats for a large set of processes, such that they can be utilized for a variety of analyses, is therefore advantageous.

In this paper we propose a framework, which we call \texttt{pineline}, that aims to generate theory predictions by 1) building a translation layer from a common input format to each of the different generators and 2) implementing a common output format for all of them.
This is the idea that we call \emph{industrialization}: while specific generators are sufficient for the calculation of single processes, there is no single generator that is able to calculate all processes, which are not necessarily limited to processes at the LHC, but may also include deep-inelastic scattering processes, for example.
By interfacing to multiple generators, and thus connecting them in an ``assembly line'' or ``pipeline'', we can easily run the generator best suited for a particular process, and by having a common input format we can easily perform variations, such as changing parameters for parameter scans.

The motivation for this project is fitting parton distribution functions (PDFs)~\cite{Ball:2021leu,Bailey:2020ooq,Hou:2019efy,Alekhin:2017kpj},
but the output generated by \texttt{pineline} can potentially be used in any fit or analysis that requires theory predictions.
One interesting feature of a PDF fit in this context is that a very large number of predictions go into it.
This complicates keeping track of the theory parameters used, for example.
While this is a manageable problem for a few predictions, for a complete PDF fit it is crucial to make sure that different processes make use of sets of parameters that are compatible with each other.
Keeping track of the parameters in a central place makes it then easily possible to be able to rerun predictions if we want to change (some) of those parameters, for example.
We stress that PDFs are a fundamental ingredient in any observable involving hadrons in the initial state and thus
should be controlled in all applications.

The rest of the paper is organised as follows: in \cref{sec:principles} we outline the abstract ideas that led us in the design
of the \texttt{pineline}. In \cref{sec:technical} we give a brief technical overview of the actual implementation
leaving the details to the respective documentations. In \cref{sec:example} we give an explicit application of the
framework, before concluding in \cref{sec:outlook}.

\section{Guiding principles}
\label{sec:principles}

We aim to bring theory predictions in high-energy physics closer to the FAIR principles \cite{Wilkinson2016}
(findability, accessibility, interoperability, and reusability) to allow for sustainable and reproducible research.

\subsection{Input and output formats}

Our framework is built to generate and store theory predictions in a single format from a common set of inputs.
By making the input common across different generators we can enforce consistency in theory settings,
and, by storing them in a single format, we ensure they can be used and analyzed regardless of how they were computed originally.

To give an impression of the diversity of generators, in NNPDF4.0~\cite{Ball:2021leu} the predictions from more than ten different programs were used: \texttt{APFEL}~\cite{Bertone:2013vaa}, \texttt{DYNNLO}~\cite{Catani:2007vq,Catani:2009sm}, \texttt{FEWZ}~\cite{Gavin:2010az,Gavin:2012sy,Li:2012wna}, \texttt{Madgraph5\_aMC@NLO}~\cite{Alwall:2014hca,Frederix:2018nkq}, \texttt{MCFM}~\cite{Campbell:1999ah,Campbell:2011bn,Campbell:2015qma,Campbell:2019dru}, \texttt{Njetti}~\cite{Boughezal:2015dva,Gehrmann-DeRidder:2015wbt}, \texttt{NNLOjet}~\cite{Britzger:2019kkb}, \texttt{NLOjet++}~\cite{Nagy:2001fj}, \texttt{Top++}~\cite{Czakon:2011xx}, \texttt{Vrap}~\cite{Anastasiou:2003ds} and \texttt{SHERPA}~\cite{Gleisberg:2008ta}.
Each of these programs requires a different set of inputs and parameters to run, and even when they are similar they are provided in different formats.
To mitigate this problem we propose a layout with a global ``theory runcard'' which,
through an appropriate generator-dependent translation layer, is fed to the target program.

The output of the programs is a hadronic observable,
which means it has already been folded with non-perturbative objects, such as the PDF.\@
By standardizing the output of all generators to be an interpolation grid
we can reanalyze the same prediction in different scenarios, without requiring an (expensive) recomputation.
The evaluation of the results for different sets of PDFs becomes almost instantaneous.
As a by-product, it also facilitates parameter fits for objects that depend on those quantities.

In the context of \pdf fitting we can think of two common scenarios:
\begin{itemize}
    \item the inclusion of new data points into the fit (coming from existing or new experiments~\cite{Gao:2017yyd,Accardi:2012qut,Anderle:2021wcy})
    \item investigate the impact of theory settings (such as the reference value of the strong coupling $\alpha_s(M_Z^2)$ \cite{Forte:2020pyp}).
\end{itemize}
Both require us to (re-)compute theory predictions for a large number of data points.
To give a concrete example of the scale of the problem, let us consider again NNPDF4.0 which fits more than 4500 data points across almost 100 different datasets.
In order to match the increasing demands from the theory side we require more and more automation to avoid
time-consuming and error-prone manual processes.

The actual objects we are working with in practice are interpolation grids~\cite{Carli:2010rw,Britzger:2012bs,christopher_schwan_2023_7499507},
which store theory predictions independently of PDFs and the strong coupling.
Interfaces of them to some generators are available~\cite{DelDebbio:2013kxa,Bertone:2014zva,Carrazza:2020gss}.
Being independent of PDFs they are ideally suited for PDF fits where they have been widely adopted, but their use is not limited to this area.
Note that by re-fitting the \pdf, any observable that depends on it will change.
However, the partonic cross sections do not depend on the \pdf{}s.
By having them stored as interpolation grids, one can update all predictions without recomputing
the most computationally heavy part of the observables.

In summary, our goal is to provide a reliable and easy-to-use workflow that connects the necessary
intermediate steps and that can be scaled to any amount of data.

\subsection{Reproducibility}

A very important aspect of joining all of these different generators in a pipeline is the reproducibility of the results:
it must always be possible to trace every prediction back to its inputs, so that any result can be independently checked by a third-party, and so that the impact of the change from a base set of parameters can be calculated.
To this end, each interpolation grid and all intermediate objects contain all the (meta)data needed to recalculate them and to verify that both are compatible with each other.
In particular, this includes: the programs used, their version numbers and random seeds, the value of relevant standard model parameters, renormalization scheme choices, phase space cuts, and Monte Carlo uncertainties.
We note that many interpolation grids publicly available on hepdata~\cite{Maguire:2017ypu} and ploughshare~\cite{ploughshare} do not include this information, though
sometimes it can be inferred from the associated publications.
However, often these data are not available, making comparisons more difficult and time-consuming.
We make that metadata explicitly available in the grids and all other outputs, from which it can be reliably and easily extracted.

\subsection{Open-source software}

All the software used in this framework is open source, to facilitate its distribution, use and maintenance.
In addition to the code, also the data are available online in formats that can be analysed with open-source tools.
Specifically, we store all metadata in the widely used YAML\footnote{https://yaml.org} format while interpolation grids
are stored as PineAPPL grids, which can be interfaced to with many programming languages.

Finally, we note that this work can also be seen as a continuation of the effort already started with the publication of the NNPDF fitting code~\cite{NNPDF:2021uiq},
giving the community all necessary tools to reproduce and perform (theory) variations of NNPDF fits.

\section{Technical overview}
\label{sec:technical}

\begin{figure*}
    \centering
    \includegraphics[width=0.8\textwidth]{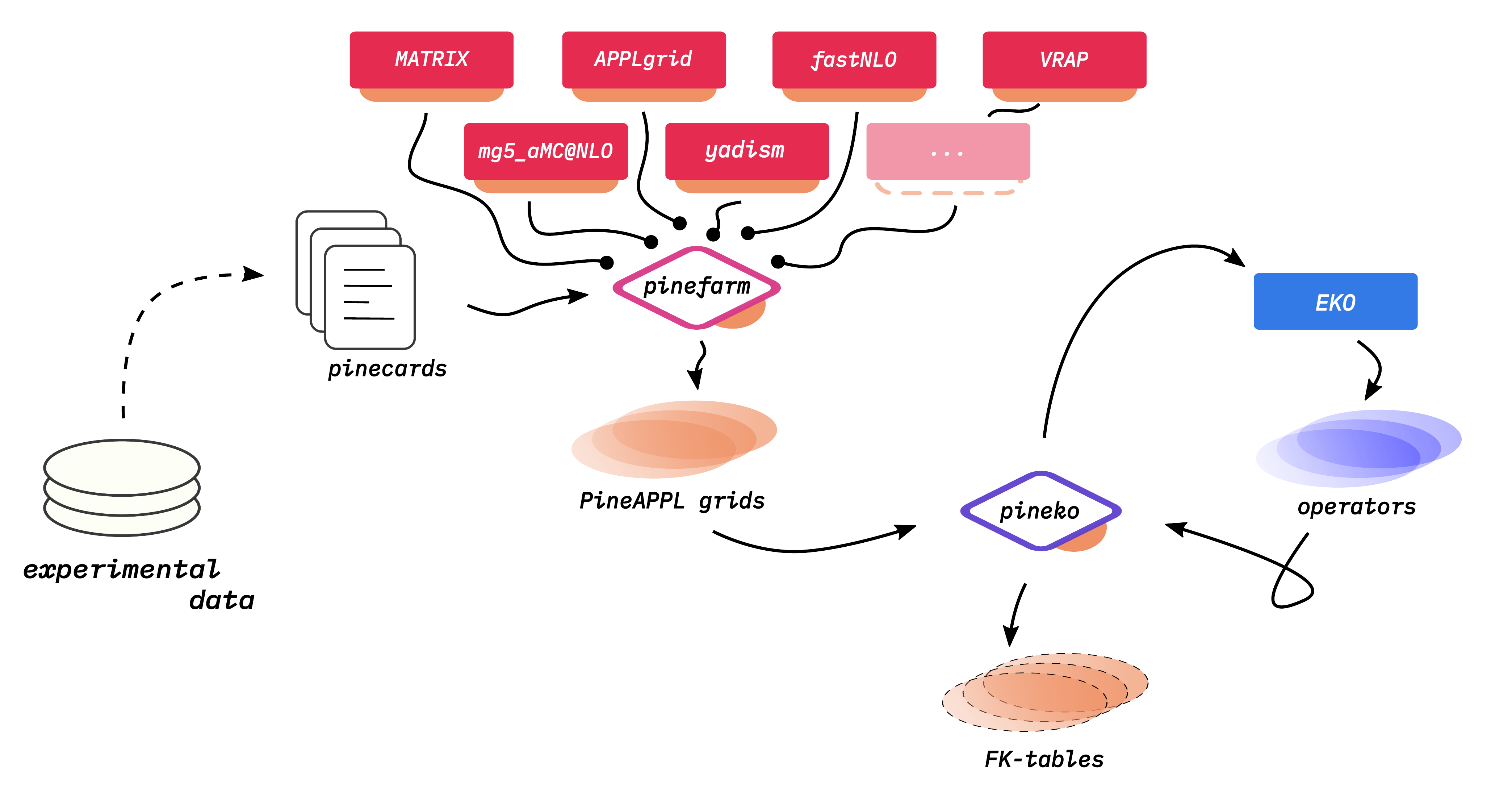}
    \caption{Flow diagram showing the overall pipeline architecture and deliverables in the case of parameter fits.
    Arrows in the picture indicate the flow of information (together with the execution order) and
    the orange insets on other elements indicate an interface to PineAPPL.
    The programs \texttt{pinefarm} and \texttt{pineko} act as interfaces between other programs and the deliverable objects,
    represented by ovals.
    These objects can be PineAPPL grids (orange) or Evolution Kernel Operators (blue).
}
    \label{fig:flow}
\end{figure*}

In the following we describe the technical implementation of the ideas highlighted above into the \texttt{pineline}.
In order to do so, it is easiest to follow the \emph{deliverables}, i.e.\ the objects that the \texttt{pineline} produces.
These are shown in \cref{fig:flow} and are the oval objects, namely 1) PineAPPL grids, 2) evolution-kernel operators (EKOs) and 3) fast-kernel (FK) tables.
PineAPPL grids, like APPLgrids and fastNLO tables, store theoretical predictions independently from their PDFs and the strong coupling.
EKOs and FK tables are tailored towards PDF fits, and translate interpolation grids to use a single factorization scale.

An extended discussion of the technical details of the various programs is beyond the scope of this paper.
We refer the interested reader instead to the relevant documentation and development repositories of each tool.

\subsection{Mathematical overview}
Let us consider
the calculation of a single observable $\sigma$, which for the sake of readability we assume to contain only a single convolution, e.g., for the case of a DIS
structure function.
The extension to more convolutions is straightforward.
\cref{eq:interpolation-grids} shows the defining property of interpolation grids,
namely how convolutions with PDFs $f_a (x, \mu_F^2)$ are performed:
\begin{equation}
\sigma = \sum_{i,j,k} \sum_a f_a (x_i, \mu_{Fj}^2) \alpha_\mathrm{s}^{n+k} (\mu_{Rj}^2) \sigma_a^{(k)} (x_i, \mu_{Fj}^2,  \mu_{Rj}^2) \,\text{.}
\label{eq:interpolation-grids}
\end{equation}
The grid itself is the set of values $\bigl\{ \sigma_a^{(k)} (x_i, \mu_{Fj}^2, \mu_{Rj}^2)
\bigr\}$ for all partons $a$ and perturbative orders $k$. 
Note that the PDFs are interpolated, and therefore evaluated at specific
momentum fractions $\{ x_i \}$ and (squared) factorization scales $\{ \mu_{Fj}^2
\}$, just as the partonic cross sections $\sigma_a$. 
For simplicity, in what follows we take the renormalization scale to be equal to the
factorization scale $\mu = \mu_\text{R}^2 = \mu_\text{F}^2$,
but the choice of scale is completely free.

The interpolation transforms the convolution integral to a sum, resulting in
the grid being a PDF-independent quantity.
In particular, the PDF is expanded over an interpolation basis, with the
expansion coefficients being the values of the PDF on some nodes.
This means the specific interpolation basis is only used in the construction of the grid, but is not relevant for the
construction of the PDF table (and so not of concern for any PDF user).

To represent interpolation grids we use the PineAPPL library~\cite{Carrazza:2020gss}.
The source code can be inspected from its repository
\begin{center}
  \url{https://github.com/NNPDF/pineappl}
\end{center}
and the associated documentation consulted at
\begin{center}
  \url{https://nnpdf.github.io/pineappl/}\,.
\end{center}

For the special case of PDF fits, interpolation grids are not the most efficient representation yet, given that the factorization dependence of the PDFs is known perturbatively and consequently not fitted.
We can therefore rewrite \cref{eq:interpolation-grids} to refer only to a single factorization scale $\mu_0$, which in PDF fits is known as the initial scale or the fitting scale:
\begin{equation}
\sigma = \sum_{i} \sum_a f_a (x_i; \mu_0^2) \operatorname{FK}_a (x_i; \mu_0^2) \text{.}
\label{eq:fk-tables}
\end{equation}
The object $\{ \operatorname{FK}_a (x_i; \mu_0^2) \}$ is known as a fast-kernel (FK)
table~\cite{NNPDF:2014otw} and is a special case of an interpolation grid that
\begin{itemize}
  \item uses a single factorization scale and
  \item contains the resummed evolution, thus combining various
    perturbative orders and therefore consuming the dependence on the strong coupling.
\end{itemize}
An FK table can be computed using evolution kernel operators (EKOs),
\begin{equation}
\operatorname{FK}_a (x_i; \mu_0^2) = \sum_{b,j,k,l} \alpha_\mathrm{s}^{n+k} (\mu_j^2) \operatorname{EKO}_{a,i}^{b,l,j} \sigma_b^{(k)} (x_l, \mu_j^2) \text{,}
\label{eq:fk}
\end{equation}
where $\operatorname{EKO}_{a,i}^{b,l,j}$ are the (linear) operators resulting from the evolution equations.
FK tables are ideally suited for PDF fits, because the time- and memory-consuming evolutions are done only once and not during the fit.

What we have gained are theoretical predictions $\{\sigma\}$, represented as FK tables, which allow us to perform convolutions with a set of one-dimensional PDFs $f_a (x; \mu_0^2)$ very efficiently.
However, the price we have to pay is that we need a set of tools that calculate all the required objects:
\begin{enumerate}
\item A numerical calculation must generate interpolation grids for each observable $\sigma$ that we want to incorporate in a fit.
\item Next, we need to calculate the EKOs, for the corresponding choices in each observable calculated previously and the choices made in the fit.
\item Finally, we need to evolve the interpolation grids using the EKOs to generate FK tables.
\end{enumerate}
In the subsequent sections we briefly review the various programs dedicated to each step.

Note that the assumption of a single scale is chosen here only to simplify the notation, but this is not
present in the actual implementation. In fact, having chosen a modularized composition
of the \texttt{pineline} allows for a simplified implementation of scale variations:
scale variation can be divided into renormalization scale variation, related to the
ultraviolet structure of the partonic matrix elements and which can thus only act on
the level of grids, and factorization scale variation related to the collinear
factorization theorem, which can either effect the split between PDFs and grids
or directly EKOs. We can use such scale variations to estimate the uncertainty
associated to the limited perturbative knowledge of perturbative QCD~\cite{NNPDF:2019ubu}.

\subsection{Generating grids: \texttt{pinefarm}}
\label{sec:pinefarm}

PineAPPL itself is physics
agnostic and therefore we need a  
parton-level generator to create and actually fill the grids.
This requires a generator to be interfaced to PineAPPL, which then sends the relevant phase-space information, i.e.\ $x, \mu_\text{F}, a, \ldots$, to PineAPPL,
which collects it in a space-efficient data structre representing $\bigl\{ \sigma_a^{(k)} (x_i, \mu_j^2) \bigr\}$ (see \cref{eq:interpolation-grids}).
Practically, this is done using an interface offered by PineAPPL, available for the programming languages C, C++, Fortran, Python and Rust.

As of now, PineAPPL has been interfaced to the following generators:
\begin{itemize}
\item Madgraph5\_aMC@NLO~\cite{Alwall:2014hca,Frederix:2018nkq} to calculate
  LHC processes, including NLO EW and QCD--EW corrections,
\item \texttt{yadism}~\cite{candido_alessandro_2022_6285149} to calculate NC
  and CC DIS processes,
\item a modified version\footnote{\url{https://github.com/NNPDF/hawaiian_vrap}}
  of Vrap~\cite{Anastasiou:2003ds} for fixed-target Drell--Yan processes, and
\item an interface to MATRIX~\cite{Grazzini:2017mhc} is in progress.
\end{itemize}
Furthermore, \texttt{PineAPPL} can convert already existing APPLgrids and
fastNLO tables into its own format using its command-line interface (CLI).
See App.~\ref{app:pineappl-import} for an example.

The program \texttt{pinefarm}, presented here for the first time,
abstracts away most of the differences of different generators.
For the generators listed above it recognizes different input
files, which specify the requested physical observable.
It also performs substitutions from a theory parameters database, and directly
runs the generators to produce predictions and collect the desired
interpolation grid.
The extension to more generators should be straightforward thanks to
the open source nature of \texttt{PineAPPL} and \texttt{pinefarm}.

The source code can be inspected from its repository
\begin{center}
  \url{https://github.com/NNPDF/pinefarm}
\end{center}
and the associated documentation inspected at
\begin{center}
  \url{https://pinefarm.readthedocs.io}\,.
\end{center}

\subsection{Generating evolution kernel operators: \eko}
\label{sec:eko}

While grids $\bigl\{ \sigma_a^{(k)} (x_i, \mu_j^2) \bigr\}$ are convoluted with PDFs evaluated at
high scales $\mu_j^2$, FK tables $\bigl\{\operatorname{FK}_{a} (x_i; \mu_0^2) \bigr\}$ are convoluted with PDFs evaluated
at the fitting scale $\mu_0^2$ reducing the dimensionality thus to just two dimensions for DIS observables (parton flavor index and momentum fraction) and four for hadronic observables.
This reduction is possible because the scale dependence of PDFs is given by the DGLAP equation~\cite{Altarelli:1977zs,Gribov:1972ri,Dokshitzer:1977sg}.

The software package \eko\cite{candido_alessandro_2022_6340153,Candido:2022tld} has been developed to solve these equations in terms of
evolution kernel operators (EKOs):
\begin{equation}
    f_b(x_l, \mu_j^2) = \sum_i \sum_a \operatorname{EKO}_{a,i}^{b,l,j} f_a(x_i; \mu_0^2)
\end{equation}
In contrast to similar programs \cite{Vogt:2004ns,Botje:2010ay,Bertone:2013vaa,Bertone:2017gds} \eko
focuses specifically on the direct computation of the operator
which allows the described pipeline to use them to produce FK tables.
Since the operator itself is PDF independent it allows also to reuse
existing operators just like reusable tools in the theory factory.

The source code can be inspected from its repository
\begin{center}
  \url{https://github.com/NNPDF/eko}
\end{center}
and the associated documentation is available at
\begin{center}
  \url{https://eko.readthedocs.io}\,.
\end{center}

\subsection{Generating FK tables: \texttt{pineko}}
\label{sec:pineko}

Interpolation grids and EKOs are joined together in \texttt{pineko}, presented here for the first time, to produce FK tables according to \cref{eq:fk}.
Specifically, \texttt{pineko} has to extract the relevant information from a grid and a theory runcard (containing all the relevant theory parameters) and then pick or, if it has not been calculated yet, compute
the required EKO as described in \cref{sec:eko}.
Once the EKO is computed, \texttt{pineko} loads the grid and evolves it using the EKO to produce the final FK table.

Since \cref{eq:fk-tables} is a special case of
\cref{eq:interpolation-grids}, PineAPPL can also represent FK tables in the same
format.
This serves an important purpose: at any point in the pipeline, a theory
prediction, whether it is an interpolation grid or an FK table, whether it was
created using a Monte Carlo generator or converted from other interpolation grids, is
always a PineAPPL grid.
Therefore, the same tools can be used on all of them.

The separation of the computation of the EKO and its convolution with the grid is
convenient from a computational point of view.
To illustrate the problem this separation solves, consider two possible scenarios:
\begin{itemize}
    \item studies on the variation of $\alpha_\mathrm{s} (M_\mathrm{Z})$ \cite{Forte:2020pyp} which require
          only the recalculation of EKOs, but not the grids (Note that in \cref{eq:interpolation-grids}
          the strong coupling is factored out)
    \item studies on the variation of $M_\mathrm{W}$ which require only the recalculation of grids, but not the EKOs.
\end{itemize}

The source code can be inspected from its repository
\begin{center}
  \url{https://github.com/NNPDF/pineko}
\end{center}
and the associated documentation inspected at
\begin{center}
  \url{https://pineko.readthedocs.io}\,.
\end{center}

\section{Application: K-factors vs.\ exact predictions}
\label{sec:example}

As an application of the previously presented tools,
we have integrated Vrap~\cite{Anastasiou:2003ds} into \texttt{pinefarm} and interfaced it to \texttt{PineAPPL} to produce FK tables for fixed-target Drell--Yan observables (FTDY) with up to
next-to-next-to-leading order (NNLO) precision in the strong coupling.
The step-by-step guide for the implementation of these results with the latest version of the \texttt{pineline} is documented at
\begin{center}
  \url{https://nnpdf.github.io/pineline/examples/vrap}
\end{center}
where only the last step is specific to the NNPDF framework.

In the following we use the framework presented in this paper and the steps outlined in the tutorial above
to produce fits similar to NNPDF4.0~\cite{Ball:2021leu}, which however differ in their treatment of predictions for the FTDY datasets: E605~\cite{Moreno:1990sf}, E866~\cite{Webb:2003ps,Towell:2001nh} and SeaQuest~\cite{Dove:2021ejl}.

\begin{figure}
    \includegraphics[width=0.5\textwidth]{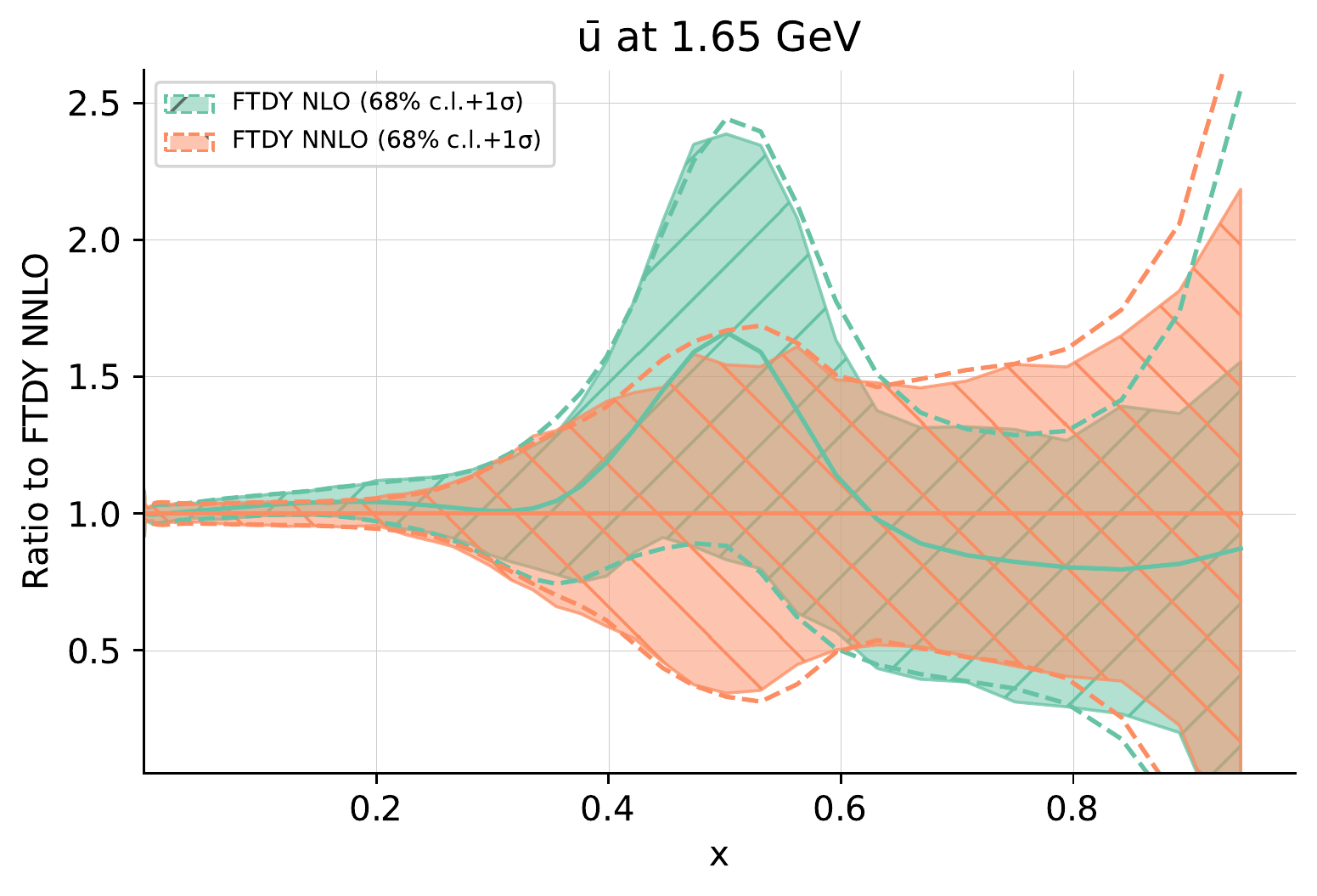}
    \includegraphics[width=0.5\textwidth]{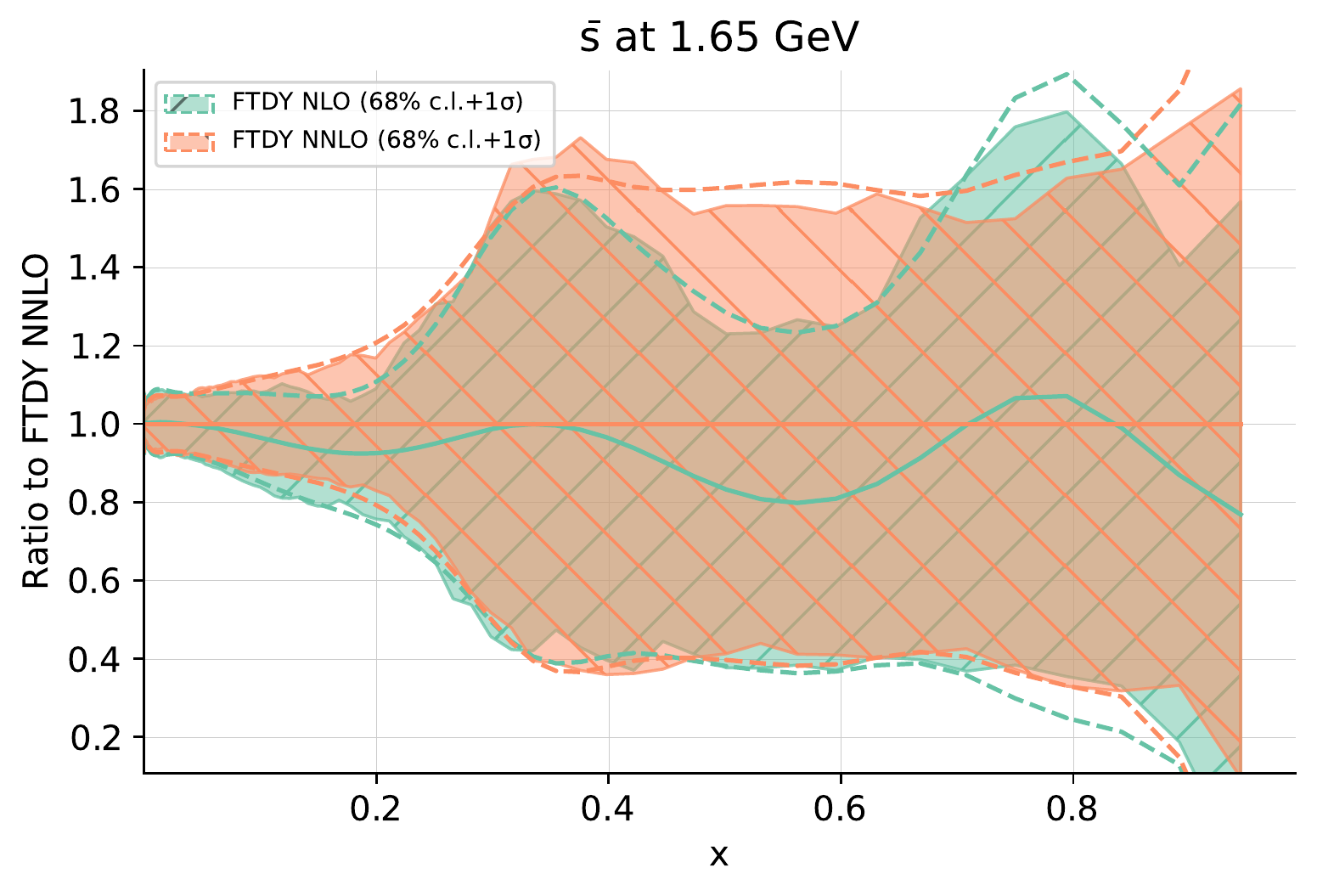}
    \caption{Comparison of PDF fits with and without NNLO contributions for FTDY in the determination. In both cases all other datasets are included at NNLO, the only difference between them is the exact NNLO contribution for FTDY.}
    \label{fig:withwithout}
\end{figure}

In particular, we change these predictions
\begin{enumerate}
\item to include \emph{only NLO},
\item to include NNLO \emph{approximately} as K-factors (as in NNPDF4.0) and finally
\item to include NNLO \emph{exactly} by using interpolation grids.
\end{enumerate}
We note that the bulk of the hadron--hadron collider data (in particular all Drell--Yan Z and W production at the LHC) in all PDF fits are still limited to NNLO K-factors.
K-factors are known to suffer from accidental cancellations between different partonic channels~\cite{Duhr:2020seh} and therefore they should be replaced by interpolation grids to produce a truly NNLO-accurate PDF fit.
However, their use is widespread when studying complex observables for which the computation of exact NNLO prediction as a grid might be very difficult, computationally expensive, or simply not publicly available.

\begin{figure}
    \includegraphics[width=0.5\textwidth]{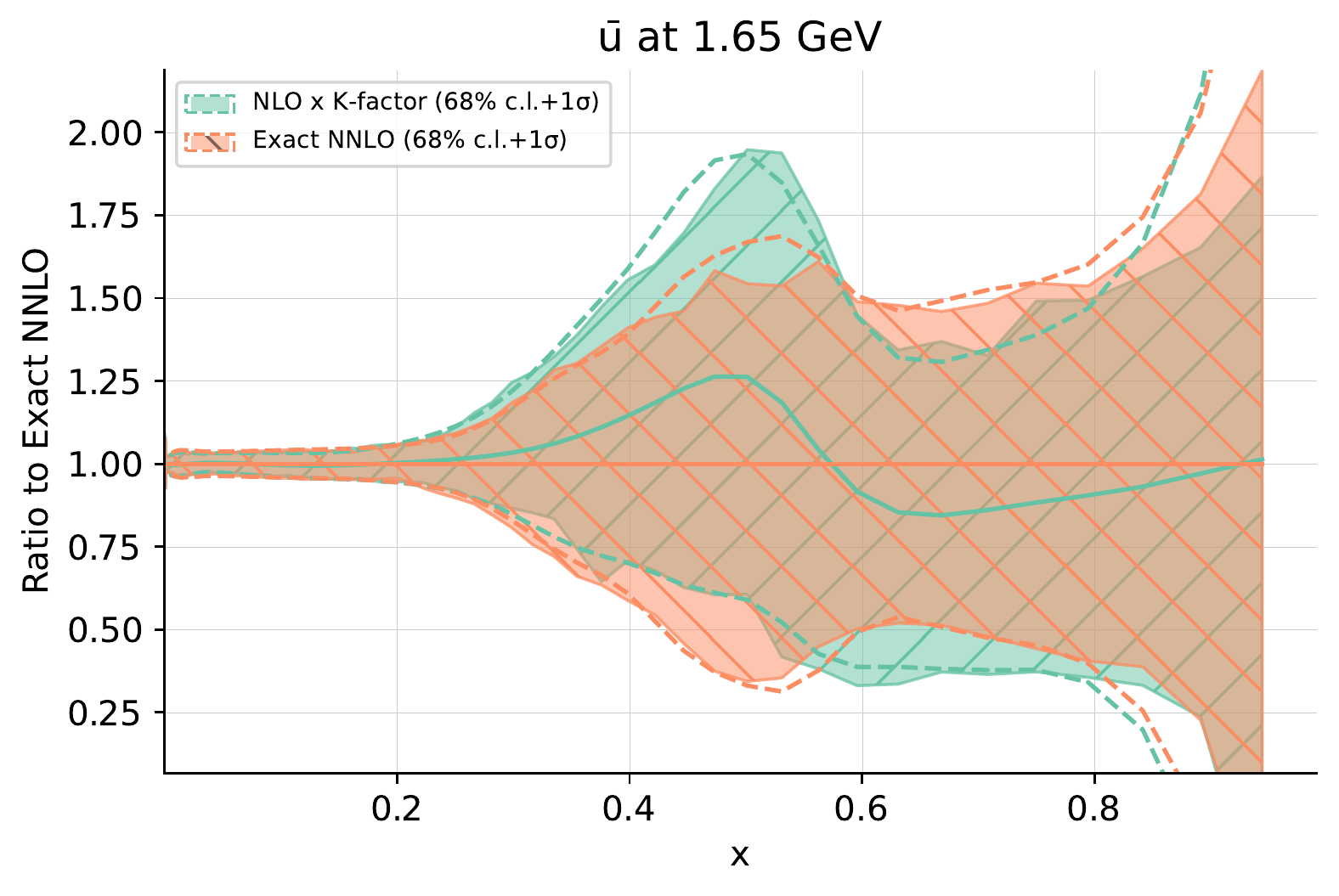}
    \includegraphics[width=0.5\textwidth]{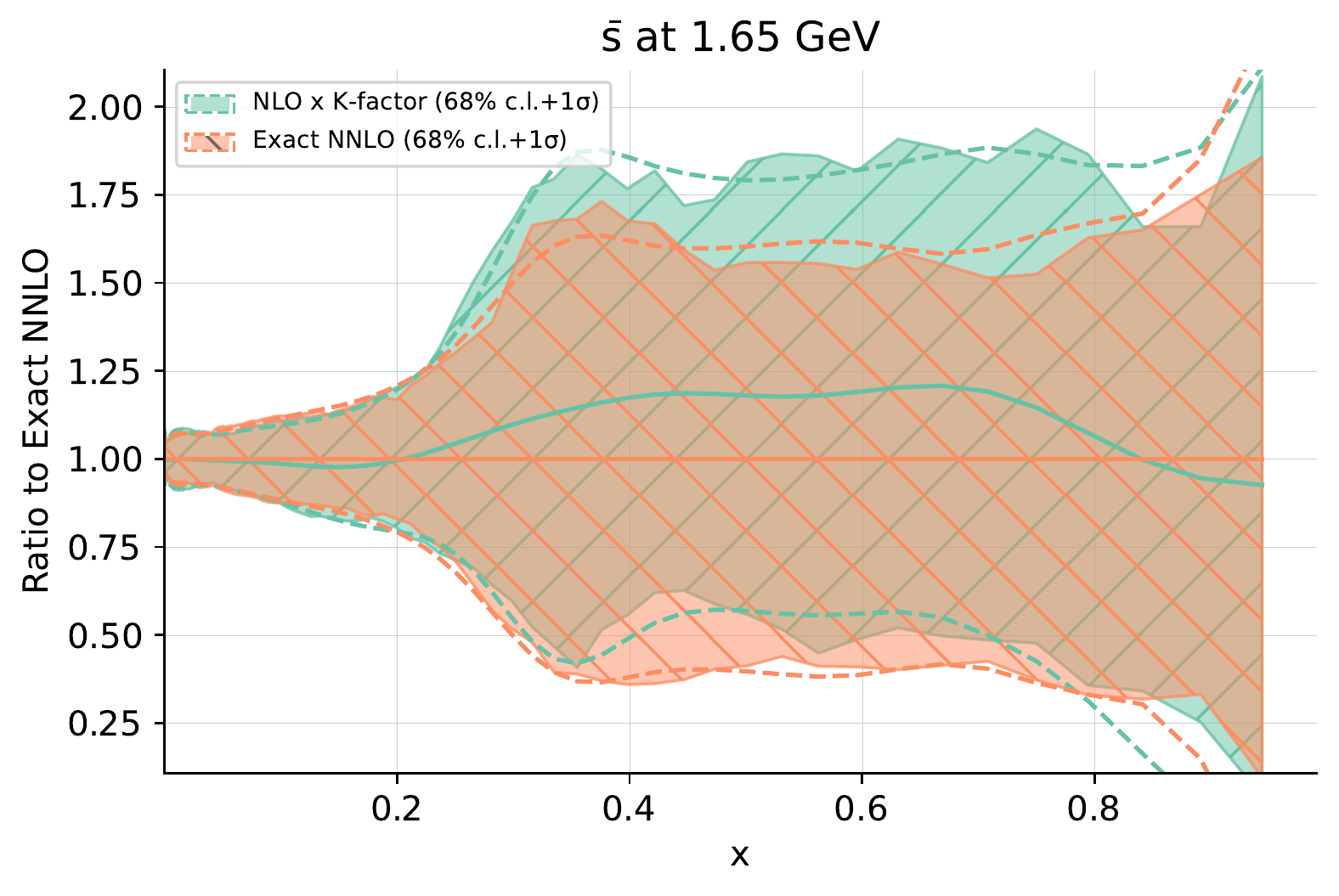}
    \caption{Comparison of PDF fits in which the FTDY datasets are included up to NNLO, including the exact predictions in the FK tables up to NNLO (orange) or up to NLO with K-factors (green). The orange fit corresponds to that of \cref{fig:withwithout}.}
    \label{fig:kfactor}
\end{figure}

\cref{fig:withwithout} shows the result of a fit including FTDY datasets only at NLO QCD (green), normalized to the results of a fit with exact NNLO QCD predictions (orange).
In \cref{fig:kfactor} we address the impact of including the NNLO contribution to the predictions in two different ways: exactly at NNLO (orange) and approximated by multiplying the NLO results by a bin-dependent K-factor (green).

\begin{figure}
    \includegraphics[width=0.5\textwidth]{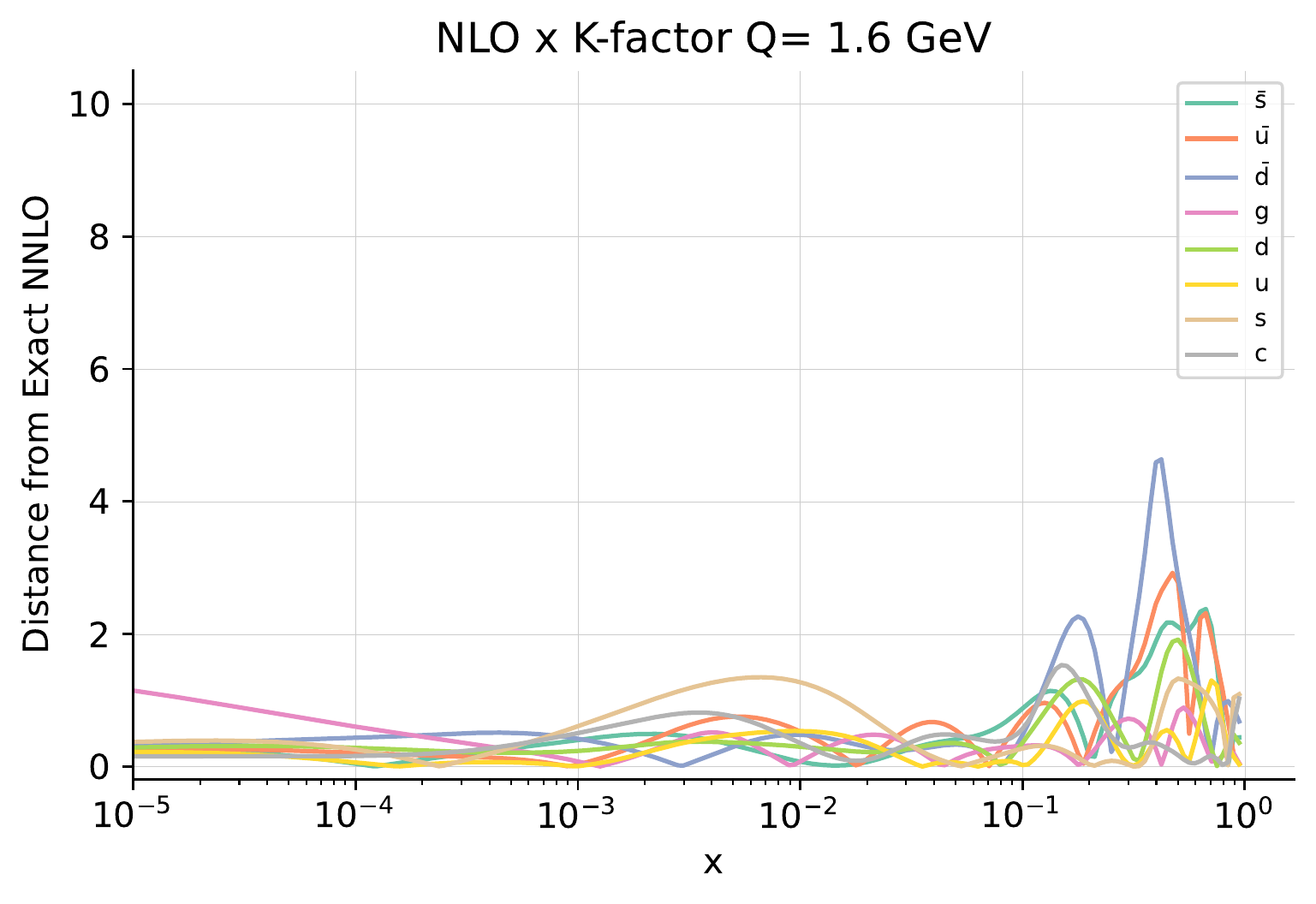}
    \caption{Distance plots between using the exact NNLO calculation and the K-factors as computed per Eq.~(48) of Ref.~\cite{Ball:2014uwa}.
    A distance of 10 units corresponds in this case to a $1\sigma$ difference between the two PDF sets.}
    \label{fig:distances}
\end{figure}

In the particular case of FTDY we note already in \cref{fig:withwithout} that the effect of NNLO corrections is constrained to a small portion of the PDF space.
In \cref{fig:kfactor} we can see that the effect of performing a fit with K-factors does move the fit in the direction expected from \cref{fig:withwithout},
but that the K-factors are not able to fully capture the nuances of the NNLO contribution.
A similar behavior is shown by the plot of the $\bar{s}$ PDF in the same figures.
These contributions are however compatible within uncertainties and the impact of using the K-factor approximation in this case is negligible.
The quantitive difference between PDFs fitted from exact NNLO contributions or the K-factors is shown in \cref{fig:distances}; the difference is never significant and stay well below half a $\sigma$.

This is just one example of a phenomenological study facilitated by the framework presented in this paper.
From a single run of Vrap we have been able to extract NLO, NLO$\times$K-factor and NNLO (QCD) predictions.
All of these predictions have been evolved to FK tables using the same NNLO EKOs, producing three
different FK tables for three different fits.
On the \texttt{pineline} site (\url{https://nnpdf.github.io/pineline}) the reader can find a tutorial for the reproduction of these results.

\section{Conclusions and outlook}
\label{sec:outlook}

In this paper we described \texttt{pineline}, which is a collection of tools that includes \texttt{pinefarm} and \texttt{pineko}.
The program \texttt{pinefarm} uses existing generators like \texttt{vrap},
\texttt{yadism} or \texttt{Madgraph5\_aMC@NLO} to generate PineAPPL interpolation grids, which in turn can be converted to FK tables with \texttt{pineko}, which uses \texttt{EKO}.
The produced objects are PineAPPL grids and store theory predictions independently from their PDFs, so that convolutions with arbitrary PDFs can be done near instantaneously after generation.
The grids are useful for phenomenological studies, and we have shown an application in which we estimate the effect of replacing NNLO QCD K-factors by the exact calculation.

We want to highlight that the PineAPPL grids produced with \texttt{pinefarm}
can be used by any external tool which includes an interface to said grids.
One such example that has recently implemented this interface is the xFitter framework~\cite{Alekhin:2014irh,Bertone:2017tig,xFitter:2022zjb,xFitter:web}.
In Ref.~\cite{Cruz-Martinez:2023sdv}, the authors present a phenomenological study making use of the software and tools presented in this paper for PDF fits with both the NNPDF and xFitter frameworks.

Below we name a few more applications for which we expect this framework to be beneficial due to the possibility of performing systematic studies of the effect of theory settings and theory predictions in PDF studies:
\begin{itemize}
  \item we need to consistently account for theory uncertainties~\cite{NNPDF:2019ubu}, coming either from the hard scattering
  process or the PDF evolution, and propagate these additional constraints into the final PDF delivery.
  \item Furthermore, it seems necessary to increase the perturbative order to next-to-next-to-next-to leading order (N3LO)~\cite{Caola:2022ayt,McGowan:2022nag}
  to match the experimental precision.
  \item Finally, we need to consider the interplay of QCD and QED~\cite{Xie:2021equ,Cridge:2021pxm,Bertone:2017bme} and eventually
  consider EW corrections in a PDF determination.
\end{itemize}
In addition, the understanding of the impact of PDF uncertainties on beyond standard model searches~\cite{Ball:2022qtp} is fundamental
in the hunt for new physics searches.

The framework is not restricted to the case of unpolarized proton PDF determination, but can already be applied to the extraction
of other factorizable objects.
Specifically, the extraction of transverse-momentum dependent PDFs~\cite{Amoroso:2022eow,Bacchetta:2022awv,Cerutti:2022lmb}
as well as the extraction of fragmentation functions~\cite{AbdulKhalek:2022laj} can be facilitated with the interpolation grids produced by this pipeline.
With the advent of the EIC projects~\cite{Accardi:2012qut,Anderle:2021wcy}
the refined determination of nuclear and polarized PDF~\cite{Ethier:2020way} will also become available.

The framework also provides a standardized way to compare theory setting in different PDF groups, and allows an easy benchmark between the respective
settings~\cite{PDF4LHCWorkingGroup:2022cjn}.

\section*{Acknowledgments}
We thank S.\ Forte and J.\ Rojo for careful reading of the manuscript and useful comments.
We also thank the members of the NNPDF collaboration for useful and insightful discussions at the various stages of this project.
A.C.\ and F.H.\ are supported by the European Research Council under the
European Union's Horizon 2020 research and innovation Programme (grant
agreement number 740006).
C.S.\ is supported by the German Research Foundation (DFG) under reference
number DE 623/6-2.

\appendix

\section{PineAPPL command-line interface (CLI) examples}
\label{app:pineappl-cli-examples}

In this section we give an overview of what can be done with the generated interpolation grids/FK tables from pineline with the PineAPPL's command-line interface (CLI) \texttt{pineappl}.
See
\begin{center}
\url{https://nnpdf.github.io/pineappl/}
\end{center}
for instructions on how to install it, and for more comprehensive documentation and tutorials.
A dedicated page for the CLI can be found at
\begin{center}
  \url{https://nnpdf.github.io/pineappl/docs/cli-tutorial.html}.
\end{center}

\subsection{Importing APPLgrids and fastNLO tables}
\label{app:pineappl-import}

The \texttt{import} subcommand of \texttt{pineappl} is able to convert existing APPLgrids and fastNLO tables into PineAPPL's format (\texttt{pineappl.lz4}).
An example of the conversion of a flexible-scale fastNLO table, generated by the APPLfast project~\cite{Britzger:2019kkb} for jet production at HERA~\cite{H1:2016goa} and downloaded from Ploughshare~\cite{ploughshare}, is shown in \cref{fig:pineappl-import}.

\begin{figure}
\begin{lstlisting}
$ pineappl import applfast-h1-incjets-fnlo-arxiv-1611.03421-xsec000.tab.gz import.pineappl.lz4 CT18NNLO
b PineAPPL fastNLO  rel. diff  svmaxreldiff
-+--------+--------+----------+------------
0  1.125e3  1.125e3  2.665e-15    7.883e-15
1  3.822e2  3.822e2  1.110e-15    8.216e-15
2  8.829e1  8.829e1  3.331e-15    5.773e-15
3  1.586e1  1.586e1 -3.997e-15    1.277e-14
4  2.725e0  2.725e0  4.663e-15    9.215e-15
5 4.848e-1 4.848e-1 -2.665e-15    1.310e-14
\end{lstlisting}
\caption{
Conversion of a flexible-scale fastNLO table into a PineAPPL grid.
The output shows five columns: 1) the bin index from 0 to 5, 2) the result after the conversion, using the given PDF \texttt{CT18NNLO}, 3) the result using the PDF calculated from the fastNLO table, 4) the relative difference between the two previous results and finally 5) the maximum relative difference between between the two results which are scale varied.
The relative differences come from differences in the fiftenth digit, as expected from double-precision results.
}
\label{fig:pineappl-import}
\end{figure}

\subsection{Convoluting grids with PDF sets}

Convolution of grids with a single or more PDF sets is done with \texttt{convolute}, as shown in the example in \cref{fig:pineappl-convolute}.
The interpolation grid corresponds to a measurement of the cross section of a single antilepton differentially in its pseudorapidity at LHCb~\cite{LHCb:2015okr}.

\begin{figure}
\begin{lstlisting}
$ wget https://github.com/NNPDF/pineapplgrids/raw/master/LHCB_WP_7TEV.pineappl.lz4
$ pineappl convolute LHCB_WP_7TEV.pineappl.lz4 NNPDF40_nnlo_as_01180 CT18NNLO
b   etal    dsig/detal      CT18NNLO
     []        [pb]         [pb]     [%]
-+----+----+-----------+-----------+-----
0    2 2.25 7.8847492e2 7.7526895e2 -1.67
1 2.25  2.5 7.2061556e2 7.1092145e2 -1.35
2  2.5 2.75 6.2526363e2 6.1876958e2 -1.04
3 2.75    3 5.0385884e2 5.0017809e2 -0.73
4    3 3.25 3.7400170e2 3.7228440e2 -0.46
5 3.25  3.5 2.5300890e2 2.5236943e2 -0.25
6  3.5    4 1.1909464e2 1.1857770e2 -0.43
7    4  4.5 2.9004607e1 2.7740964e1 -4.36
\end{lstlisting}
\caption{
Convolution of a PineAPPL grid with two PDF sets.
This shows the following columns: 1) the bin index from 0 to 7, 2) the left and 3) the right limit of each bin and 4) the prediction for each bin using the first PDF set.
For each additional PDF set given the 5) absolute numbers are shown and 6) the relative differences to the first PDF set.
Using metadata the first (and only) bin dimension is denoted as `etal', the pseudorapidity of the antilepton, which does not have a unit, shown by the empty square-brackets.
The prediction is shown as `dsig/detal`, so the cross section differentially in the pseudorapidity of the antilepton, in units of picobarn.
}
\label{fig:pineappl-convolute}
\end{figure}

\subsection{Other subcommands}

PineAPPL's CLI \texttt{pineappl} offers many more subcommands that allow to convolute interpolation grids with PDFs in different ways: apart from \texttt{convolute} it can show the results separately by perturbative orders (\texttt{orders}), to read off the impact of higher-order corrections, and separately by channel (\texttt{channel}) to read off the hierarchy of which partonic initial states are most important.
PDF and scale uncertainties can be calculated by \texttt{uncert}, and \texttt{pull} can be used to understand the differences of two PDFs: how large and significant they are and from which channels they come from.
Two interpolation grids can be compared against each other using \texttt{diff}, \texttt{merge} combines two or more grids by combining them correctly and \texttt{plot} writes a Python plot script, which graphically display most of the information given numerically above.
Finally, the subcommands \texttt{read} and \texttt{write} can be used to read out data and metadata from a grid and perform operations and set metadata, respectively.


\bibliographystyle{elsarticle-num}
\bibliography{./my}

\end{document}